# Consequences of Utilizing a Redox-Active Polymeric Binder in Li-ion Batteries

Sathish Rajendran, Haoyu Liu, Stephen E. Trask, Baris Key, Andrew N. Jansen, and Marco-T. F. Rodrigues[*]

Chemical Sciences and Engineering Division, Argonne National Laboratory, Lemont, Illinois, USA.

**Abstract:**

Development of new polymeric binders can help enable the use of silicon-rich anodes in Li-ion batteries, by providing stronger adhesion to the active material particles. The compositional features that improve interfacial interactions and mechanical properties can often impart electronic conductivity and redox activity to these polymers, which are generally seen as beneficial to cell performance. Alternatively, it is also possible that the addition of charge-transferring centers to the electrode can accelerate cell degradation. Here, we use an aromatic polyimide (~320 mAh/g of reversible capacity) to explore how a redox-active conductive polymer can affect cell performance. We demonstrate that the lithiated polymer is less stable than the traditional binders upon storage, leading to increased rates of calendar aging. Furthermore, we show that the adhesion properties of the polymer deteriorate upon repeated cycling, to an extent that is proportional to the degree of delithiation of the binder. More critically, we show that progressive degradation of the redox behavior of the polymer leads to the release of "extra" Li$^+$ into the cell, which can give the false perception of good performance even under conditions of poor stability. Our work suggests that redox-active conductive binders can sometimes be detrimental to cell performance, and that works evaluating new polymers must include careful experimental validation under realistic conditions.

## 1. Introduction

Li-ion batteries play a vital role in electric vehicles, portable electronics, and grid energy storage due to their higher energy density and cycle life when compared to other energy storage technologies.[1] Despite the improvements in the energy metrics over the past two decades, there is still a need of enabling cost-effective improvements. Hence, the energy density of the Li-ion battery technology needs to be drastically improved to meet the current and future demands.[2] One of the promising methods is to replace the graphite anode (theoretical capacity 372 mAh/g) with silicon ($Li_{3.75}Si$; theoretical capacity 3579 mAh/g) that boosts the anode capacity by nearly 10 times.[3] However, silicon undergoes large changes in its bonding environment upon lithiation that results in drastic volume changes (>300%).[4] These dimensional changes tend to isolate the active material in the electrode, leading to rapid capacity losses and resulting in poor cycle life of the battery.[5] An important reason behind the isolation of active material is the weak Van-der Walls forces that operate between Si particles and the widely used polyvinylidene fluoride (PVDF) binder that is not quite adequate to withstand drastic volume expansions.[6] To overcome this, various polymeric binders with different functional groups were proposed that increase the binding ability due to the bonding between the functional groups and Si particles.[6a] Some of the widely explored polymeric binders are polyacrylic acid (PAA),[7] polyvinyl acrylate (PVA),[8] sodium alginate,[9] and carboxy methyl cellulose (CMC).[10] Although these polymers exhibit superior performance than the conventional PVDF binder, they fail to relieve the stresses developed in the electrode because of their rigidity and brittleness, which may contribute to the development of cracks over the electrode surface during cycling.[11] A superior binder with suitable mechanical properties and high binding ability is required for Si anodes. Recent works have explored polymeric materials made up of aromatic, $\pi$-conjugated units with various functional groups that could offer superior adhesion between the binder, the Si particles and the current collector, and that was able to tolerate the volume expansion in Si anodes.[12]

Apart from having superior mechanical and binding properties, polymers containing $\pi$-conjugated units present another interesting feature. These materials undergo electrochemical doping with cations and anions at suitable environments and electrochemical potentials,[13] which leads to the formation of charged species known as soliton, polaron, and bipolaron.[14] Soliton formation takes place when a double bond is broken to form a single charge or two charged species on the polymer chain. Addition or removal of a single electron to form a negative or positive charged state is called a polaron, which has ±1 spin state. When a second electron is removed or added, it results in the formation of spinless bipolarons.[14b] Formation of such charged species transforms the electronically insulating phase of the polymer into an electronically conducting phase by injection of states from the conduction band into the valence band, with conductivity values approaching that of metals.

In a Li-ion battery, the anode will experience very low electrical potentials in an inert and moisture-free environment, which is perfectly suitable for the $\pi$-conjugated polymeric binders with functional groups to undergo n-doping electrochemically. In this case, $Li^+$ acts as the dopant (a common n-dopant) and the extent of doping/dedoping depends on the cut-off potentials being used.[15] The reversible redox capacity of these materials can be as high as 1500 mAh/g depending on the molecular structure of the polymer backbone chain.[13a, 16] Utilizing such materials as binders could be advantageous because of their redox activity and electronic conductivity that can help in boosting the capacity and fast charging ability of the electrode.

Nevertheless, it is also possible that deployment of redox-active conductive polymers could negatively affect cell performance. Li-ion batteries are metastable devices in which durability is achieved thanks to the formation of a passivation layer (the solid electrolyte interphase, or SEI) at the anode surface.[17] Having additional conducting surfaces and redox centers in the electrode could introduce hotspots of reactivity, threatening to disrupt a fine stability balance that has been achieved after decades of optimization in commercial devices. With that in mind, the present work uses an aromatic polyimide-based copolymer (PI, Figure 1a) as a model system to elucidate the consequences of employing such polymeric binders in Li-ion batteries. We report that the PI binder presents high capacity that is accessed with huge hysteresis, minimizing the binder contributions to reversible cell capacity, and decreasing the initial coulombic efficiency. We also show that the redox-active PI binder is less stable than PVDF during extended storage, accelerating the calendar aging of the cell. Finally, we demonstrate that this binder can interfere with the diagnostic of aging through a mechanism that releases extra $Li^+$ into the cell, making the measured capacity fade to be smaller than the true extent of aging experienced by the cell. Our study highlights the practical issues that may arise when using functional polymer binders and indicate that novel compounds must be thoroughly analyzed before being presented as a viable option for usage in Li-ion batteries.

2. Experimental Methods

*Electrode Preparation:* The polyimide (PI) binder was purchased directly from the manufacturer. PI powder was dissolved in N-methyl pyrrolidone (NMP; Sigma Aldrich) to make a 20 wt. % solution by mixing in a rotary mixer for 24 hours at 50-75 rpm. Titanium nitride (TiN) with 99.5 % purity was purchased from Alfa Aesar, C-45 conducting carbon from Timcal, Graphite from Hitachi (MagE3), and polyvinylidene difluoride (PVDF) from Kureha. The precursors in suitable weight ratios were mixed with an adequate amount of NMP in a Thinky mixer at 2000 rpm for 4 minutes and 2200 rpm for 30 seconds to form a viscous slurry. The slurry was coated on a copper foil using the doctor blade method, and the resulting electrode was dried overnight at 75 °C in a hot air oven. NMC811 and Graphite electrodes with PVDF binder were fabricated by the Cell Analysis, Modeling, and Prototyping (CAMP) facility at ANL. All electrodes were calendered to the indicated porosity. Electrodes with PI further underwent a curing step at 350 °C under vacuum for 1 hour to enhance the cross-linking within the polymer chains. Electrodes containing PVDF were dried at 120 °C under vacuum overnight prior to use.

The electrode composition and porosities are stated below:

- TiN-PI electrode: 95% TiN and 5% PI; Porosity: 38.1%; Coating mass loading: 5.1 mg/cm$^2$.
- Graphite-PVDF electrode: 91.83% Graphite, 0.17% oxalic acid, 2% C-45, 6% PVDF; Porosity: 30.3 %; Coating mass loading: 6.35 mg/cm$^2$.
- Graphite-PI electrode: 92% Graphite, 2% C-45, 6% PI; Porosity: 32.2%; Coating mass loading: 5.35 mg/cm$^2$.
- NMC 811 electrode (used to pair with Graphite-PVDF pouch cells): 95.70% NMC811 (LiNi$_{0.8}$Mn$_{0.1}$Co$_{0.1}$O$_2$), 0.05% SWCNT (single wall carbon nanotubes), 2% C-45, 2.25% PVDF; Porosity: 38.8%; Coating mass loading: 8.17 mg/cm$^2$.
- NMC811 electrode (used to pair with Graphite-PI coin/pouch cells and Graphite-PVDF coin cells): 95.70% NMC811, 0.05% SWCNT, 2% C-45, 2.25% PVDF; Porosity: 37.4%; Coating mass loading: 9.67 mg/cm$^2$.
- TiN-PVDF electrode: 95% TiN and 5% PVDF; Porosity: 48.6%; Coating mass loading: 3.65 mg/cm$^2$.

*Cell Fabrication:* All electrochemical cells were fabricated in a dry room with a dew point lower than -45 °C. Half-cells in a coin cell format were made using the coated electrode (diameter 14 mm) and Li foil (16mm; 600μm; American elements). Full-cells in a coin cell format were fabricated using 14-mm cathode and 15-mm anode discs. Single layer pouch cells (xx3450 format) were fabricated using 14.1 cm$^2$ cathodes and 14.9 cm$^2$ anodes, and they were tested under a stack pressure of 4 PSI. The electrolyte in all cells was 1.2 M LiPF$_6$ in 3:7 EC: EMC along with 3 wt.% FEC as additive; FEC is a common additive for Si cells and was used here to explore the behavior of the PI under realistic conditions. A total of 60 μL of electrolyte was added to half-cells (Li metal anode), 40 μL to full-cells (coin cells) and 350 μL to pouch cells. These volumes constitute at least four times the total pore volume of electrodes and separator in the cells. Celgard 2500 was used as the separator for all the electrochemical cells.

*Electrochemical Testing:* Electrochemical testing of the fabricated cells was carried out using a MACCOR cycler. All testing was carried out at 30 °C; for calendar aging experiments, static aging (but not cycling) was performed at 45 °C. The reproducibility of the data was ensured by assembling multiple cells of each type. In many cases, only representative data will be shown. When relevant, we will present average values and error bars expressing the standard deviation of the data.

TiN-PI electrodes were tested in half-cells (vs. Li metal) with 5 mV as the lower cut-off and either 2 V or 600 mV as the upper cut-off at C/5 rate. The cut-off voltage was not increased beyond 2 V to ensure the Cu corrosion would not impact the electron transfer from the material and resulted in some capacity contribution. Most of the charge-discharge measurements were carried out in the galvanostatic mode unless specified. Some electrochemical half-cell testing included a potentiostatic step at the end of the galvanostatic step. The potentiostatic step involved holding the cut-off voltage until the current reached C/50 value with an upper limitation of C/5 current. The specific capacity calculations were solely based on the PI binder's weight, and the theoretical reversible specific capacity of the PI binder was assumed to be 600 mAh/g for the 'C' rate calculations.

NMC811 and Graphite-PI/Graphite-PVDF coin and pouch cells were tested with an upper cut-off voltage of 4.2 V and various lower cut-off voltages as specified in the results section. All full-cells were subjected to formation cycles at slower rate (C/20) before starting the long cycling tests. The formation cycles were performed with a lower cut-off of 3.2 V (PVDF) or 3.3 V (PI). Coin cells were exposed to cycle life testing, comprising aging cycles at C/2 (using the appropriate lower cutoff voltages) and two reference performance test (RPT) cycles at C/10 rate every 25 aging cycles. The RPTs used the same lower cutoff voltage for all cells to allow for direct comparison of aging trends resulting from the various testing conditions. Pouch cells were used in calendar aging tests, in which cells were stored at 45 °C for varying periods and were occasionally brought down to 30 °C for RPTs. RPT cycles were carried out at C/20 rate with 3.2 V (PVDF) or 3.3 V (PI) lower cut-off voltage and 4.2 V as the upper cut-off voltage. The capacity calculations were solely based on NMC811 active material's weight.

*Solid-state nuclear magnetic resonance (NMR) spectroscopy:* NMR spectroscopy was performed on the cycled samples. The cycled cell was opened inside an argon filled glovebox and the material from the cycled electrode was scrapped out using a scalpel blade and was packed into the 3.2 mm NMR rotor inside the glovebox. Magic-Angle-Spinning (MAS) NMR experiments were carried out using a Bruker Avance III spectrometer at 7.04 T (Larmor frequencies: 300.0 MHz for $^1$H and 116.6 MHz for $^7$Li) at a spinning rate of 20 kHz. Both $^1$H and $^7$Li spectra were collected with rotor-synchronized spin-echo pulse sequences (90° pulse length: 2.25 μs and 3.50 μs respectively). Long enough pulse delays (5 s and 15 s respectively) were given for full spin relaxation after each scan. The initial time-domain signal from free induction decay (FID) until echo top were removed to minimize distortion in the processed spectra partly due to the existence of conductive TiN. Tetramethylsilane (TMS) and 1 M LiCl were utilized as references for $^1$H and $^7$Li NMR (both at 0 ppm).

**Results and Discussion**

*2.1. Electrochemical activity of the PI binder*

A representative structure of the polyimide is shown in Figure 1a.[12a] This PI is obtained by the co-polymerization of 3,3,4,4-benzophenone-tetracarboxylic dianhydride (BTDA) with toluene diisocyanate (TDI) and methylene diphenyl diisocyanate (MDI), resulting in a structure that, in principle, would not require thermal treatment for imidization. Each repeating unit contains two imide groups and a total of five carbonyls, along with 5 aromatic groups that are responsible for the interchain π-stacking interactions. The PI is partly conjugated with the -CH$_2$- and imide groups acting as conjugation breakers. Despite the presence of conjugation breakers, the π-stacking interactions and electron hopping may provide electrical conductivity upon electrochemical doping.[18]

To study the electrochemical activity of the PI, excluding the capacity contribution from other electroactive components (e.g., conductive carbon) is a necessary step. Hence, model electrodes were prepared containing PI as both the active material and binder and using TiN as a redox-inactive conducting additive. The use of TiN for screening the activity of the binder was successfully demonstrated by Wilkes et al.[19] For reference, tests with an electrode containing solely TiN and PVDF resulted in <1 mAh/$g_{TiN}$ of reversible capacity, as shown in Figure 1b. Electrodes containing 5 wt% of PI were evaluated in half-cells between 5 mV and 2 V in three types of tests: i) electrodes were "cured" at 350 °C in vacuum for 1 hour to activate residual cross-linking sites between the PI chains, and were cycled galvanostatically at C/5 rate; ii) cured electrodes tested with a potentiostatic step at the voltage cutoffs until the current reaches C/50 rate (to remove any kinetic barrier for charge transport within the polymer); iii) uncured electrodes, dried at 120 °C in vacuum for 8 hours and tested galvanostatically at C/5 rate. All three electrodes exhibited very high capacities of at least ~1500 mAh/g during the first lithiation (Figure 1c). Curing the electrode raised this initial capacity to 1925 mAh/g, while adding a potentiostatic hold further increased this value to 2200 mAh/g. Upon prolonged cycling, the specific capacity of the PI drastically faded in the first 20 cycles and attained a stable value of about 320, 160 and 60 mAh/g for the cured electrode with potentiostatic step, cured electrode and the uncured electrode, respectively, in about 50 cycles. The performance bump gained from the thermal treatment likely stemmed from the improved charge transfer through the polymer. Additionally, the doubling in reversible capacity obtained with a potentiostatic step suggests that reaction kinetics is relatively slow for this PI. All in all, this binder presents a reversible capacity that is on par with many electrode active materials.

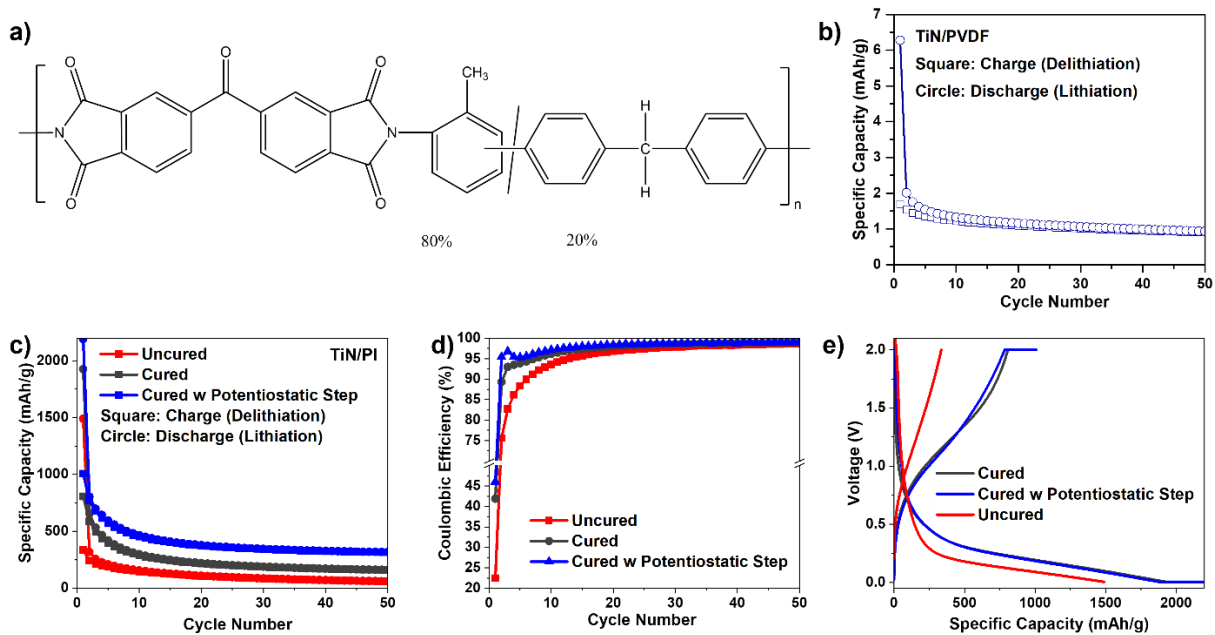

**Figure 1:** a) Structure of a model PI binder; b) Electrochemical cycling of a TiN-PVDF half-cell; c) Electrochemical cycling of half-cells made using TiN-PI electrodes at C/5 rate, and its corresponding d) Coulombic efficiency; e) First cycle voltage profile of the half-cells cycled in-between 2 V and 5 mV.

Despite the ultra-high first cycle capacity, the initial reversibility of the PI was quite poor, as indicated by coulombic efficiencies of 22%-45% (Figure 1d). Although the efficiency increased upon cycling, the average efficiency was at best 97.1 % (cured with potentiostatic hold) and only 94.3 % for the uncured electrode. Figure 1e shows the voltage profile of the first cycle for each test. The profiles were sloppy, with a significant portion of the lithiation capacity was obtained below 500 mV. This shape suggests that most of the capacity may be coming from doping/dedoping rather than from the redox activity of carbonyl groups. Carbonyls typically yield a flat voltage profile at around 1.5 V or higher depending on its environment.[20] Indeed, the capacity contributed by carbonyls would only reach ~230 mAh/g (5 electrons per repeating unit), while the measured initial capacities can reach the equivalent of 48 electrons being exchanged. The lithiation of the uncured electrodes took place at lower potentials than the cured electrodes, which may be due to the high internal resistance caused by poor ion transfer within the polymer. Voltage hysteresis was very large for the PI: after lithiation to 5 mV, meaningful delithiation capacities are only achieved >800 mV.

Various mechanisms have been proposed for the charge storage behavior of PI and similar molecules. The widely accepted mechanism is the redox activity of the functional group attached to the conjugated aromatic moiety undergoing electron transfer that leads to the delocalization of the π-electrons, resulting in solitons, polarons, and bipolarons.[14b] The possibility of the benzene ring being electrochemically active in conjugated aromatic molecules have also been proposed.[16, 21] Obrovac et al. proposed a mechanism of decomposition of the polymer during lithiation that would result in the formation of new redox active species that may exhibit a small reversible capacity.[19] To obtain more clarification on the redox activity of PI, solid-state nuclear magnetic resonance (SS-NMR) spectroscopy analysis was performed on the model TiN-PI electrodes after 30 cycles, both in the lithiated (5 mV) and the delithiated (2 V) state, as shown in Figure 2. The $^1$H NMR (Figure 2a) shows at least two distinct peaks for both samples, one at higher (7-10 ppm) and another at lower (1-4 ppm) shifts. The same set of peaks could be observed for the pristine PI solution NMR in deuterated dimethyl formamide (DMF) solvent (Figure 2b). The peaks at higher chemical shift correspond to the protons attached to the aromatic group.[22] This is because of the deshielding effect experienced when the induced current (from the aromatic ring) creates a magnetic field in the same direction as the external field. The PI contains aliphatic groups, as seen from the structure in Figure 1a, that are responsible for the shielded peaks at lower chemical shift values. After 30 cycles, the delithiated PI exhibited two peaks at 8.58 ppm and 2.97 ppm. Upon lithiation, the peak corresponding to the protons in the aromatic groups shifted towards the shielded region to 8.01 ppm, whereas the aliphatic peak at 2.97 ppm remained at the same position. Apart from these two peaks, a new shoulder appeared at 5.21 ppm that corresponds to a less aromatic (containing fewer double bonds) region. This signifies that the aromatic groups are becoming less aromatic by accommodating $Li^+$ into the PI structure. Thus, the reversible redox sites appear to be the $\pi$-electrons of the benzene rings, although the broad peaks prevent assignments to specific moieties of the repeating unit. The observation that the aromatic rings are the source of the reversible capacity agrees with previous reports that have shown through infrared spectroscopy that the carbonyl peaks of PIs disappear during the first lithiation and do not regenerate during the subsequent delithiation steps.[23] That is, carbonyls are lithiated during the first cycle and appear to remain so during the remainder of testing.

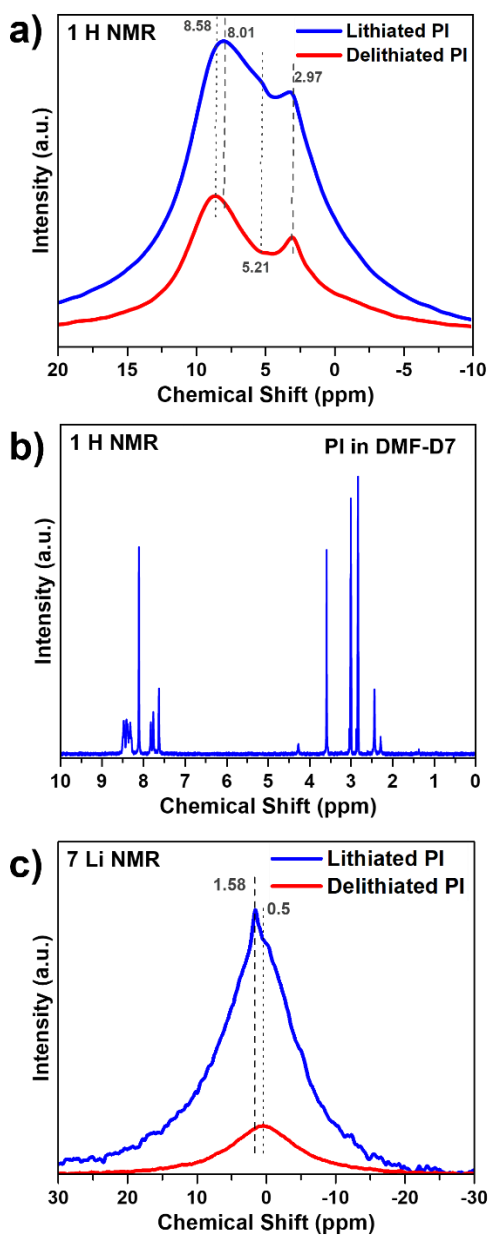

**Figure 2:** a) ¹H MAS NMR spectra on the lithiated and delithiated PI electrodes, b) ¹H NMR of PI dissolved in deuterated dimethyl formamide, and c) ⁷Li MAS NMR spectra on the cycled PI electrodes.

$^7$Li NMR for the delithiated PI shows a single peak centered at 0.5 ppm (Figure 2c). Presence of $Li^+$ in the delithiated sample indicates that $Li^+$ remains trapped within the PI structure after the 1$^{st}$ cycle. A very small contribution could also come from the residual Li salt from the electrolyte (although the sample was rinsed multiple times in dimethyl carbonate to remove salt residue before the analysis).[24] The lithiated PI exhibited an additional sharp peak at 1.58 ppm and the area under the curve was 5.4 times larger than the delithiated PI, indicating the existence of larger amounts of $Li^+$ within the PI, since the integrated peak area is directly proportional to the $Li^+$ content.[25]

In summary, the redox of the PI binder has a somewhat slow kinetics but can deliver reversible capacities of >160 mAh/g, depending on thermal treatment, cycling rate and voltage window. Furthermore, solid-state NMR measurements indicated that reversible electroactive centers exist within molecular environments that are also present in the pristine polymer, and mostly involve the aromatic rings. The voltage profile of the polymer during delithiation (Figure 1e) shows that meaningful contribution of the binder to the reversible cell capacity will only occur under relatively deep delithiation conditions. Otherwise, the polymer will irreversibly trap even the $Li^+$ that are, in principle, reversible, resulting in even lower initial coulombic efficiencies. Next, we evaluate how properties of the binder may depend on testing conditions.

*2.2. Voltage cutoff and binder degradation*

Among the many features shown in Figure 1, two are of particular interest: the PI binder exhibits i) rapid initial capacity decay and ii) negligible delithiation capacity below 600 mV. These characteristics allow us to test the individual role of lithiation and delithiation on the stability of the binder.

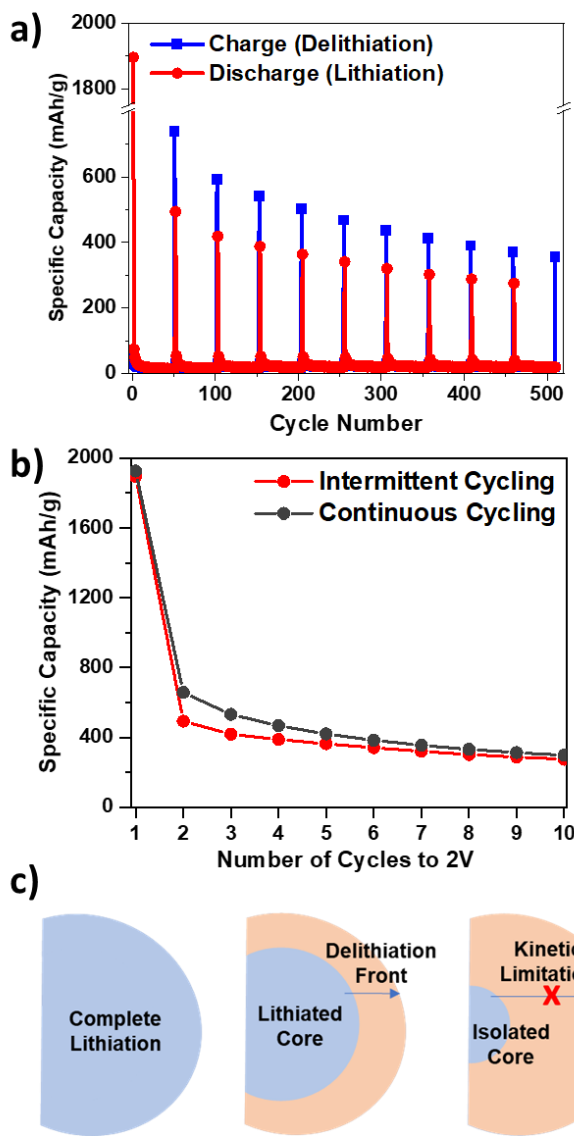

**Figure 3:** a) Electrochemical cycling of TiN-PI half-cell to 600 mV upper cut-off voltage and intermittent oxidation to 2V every 50 cycles and b) Comparison of the lithiation capacity obtained in the intermittent cycling versus a cell with continuous cycling to 2V upper cut-off voltage. c) Schematic of the proposed core-shell mechanism for capacity fade.

Figure 3a exhibits the lithiation and delithiation capacities for a TiN-PI electrode that only intermittently experienced complete delithiation. In these tests, the delithiation cutoff was limited to 600 mV, and only allowed to reach 2 V every 50 cycles without any potentiostatic hold.

Negligible capacity is exchanged during the shallow cycles: the polymer remains mostly in the lithiated state during the entire duration of the cycle. If residence in this lithiated state was the main driver of capacity fade, the full cycles (up to 2 V) should exhibit faster capacity decay than the control (the cells in which all cycles experience delithiation to 2 V without the potentiostatic hold step, Figure 1c). If instead the delithiation of the binder is when performance loss occurs, the rate of capacity fade of full cycles would match that of the control. Figure 3b clearly shows that the second assumption is correct, as the decay in both cases are nearly identical. That is, once $Li^+$ is removed from the binder, part of it can never go back. A possible hypothesis is that the relationship between capacity fade and delithiation occurs due to a "core-shell" mechanism (Figure 3c). Assume that contiguous polymer domains exist throughout the electrode, in a way that not all of the PI is in direct contact with the electrolyte. Once the outer layers of these PI domains are lithiated, the electronic conductivity will increase locally due to the formation of polarons. These conductive shells will favor electron and $Li^+$ transport to inner polymer chains, increasing the utilization of the PI binder. During delithiation, however, $Li^+$ extraction is more facile from the outer layers, which quickly return to their non-electron-conducting ground state. Once a critical thickness of delithiated polymer is achieved, $Li^+$ extraction from the lithiated domains within becomes kinetically frustrated, trapping some amount of capacity in the PI core. Existence of lithiated domains even after the electrodes are delithiated to 2 V is suggested by NMR (Figure 2c). In this case, the capacity fade shown in Figure 1 would not necessarily be due to irreversible degradation of the polymer.

There is also evidence that, to some extent, the basic adhesive properties of the polymer are changing. Figure 4 shows a series of photos collected during qualitative "scratch tests", in which different TiN-PI electrodes at various states of charge are scraped off the current collector using a scalpel. The polyimide possesses much stronger adhesion when compared to conventional PVDF binder, making it extremely difficult to delaminate the electrode material. Upon lithiation to 5 mV, the polymer remained strongly adhered to the copper, with only a superficial layer of the coating being removed (Figure 4a); this is similar to what is observed for a pristine electrode (not shown). When the electrode is fully lithiated and then delithiated to 600 mV (Figure 4b), it still retains strong adhesion to the current collector. However, further delithiation to 1 V and higher (Figure 4c-e) will lower the binding ability of the PI. At 1 V, a part of the electrode could be completely scraped out of the copper foil (Figure 4c), which became even easier after further increasing the delithiation cutoff to 1.5 V and 2 V (Figures 4d-e). Repeated cycling in between 5 mV and 2 V further facilitated the removal of the active material from the current collector (Figure 4f). In other words, it is clear that the binder loses its superior binding ability upon repeated delithiation. We were unable to identify unambiguous evidence of structural changes in the polymer that could explain this behavior, and a more detailed study in this area is beyond the scope of this work. Nonetheless, the consequences of this observation for battery testing are extremely relevant and are analyzed further.

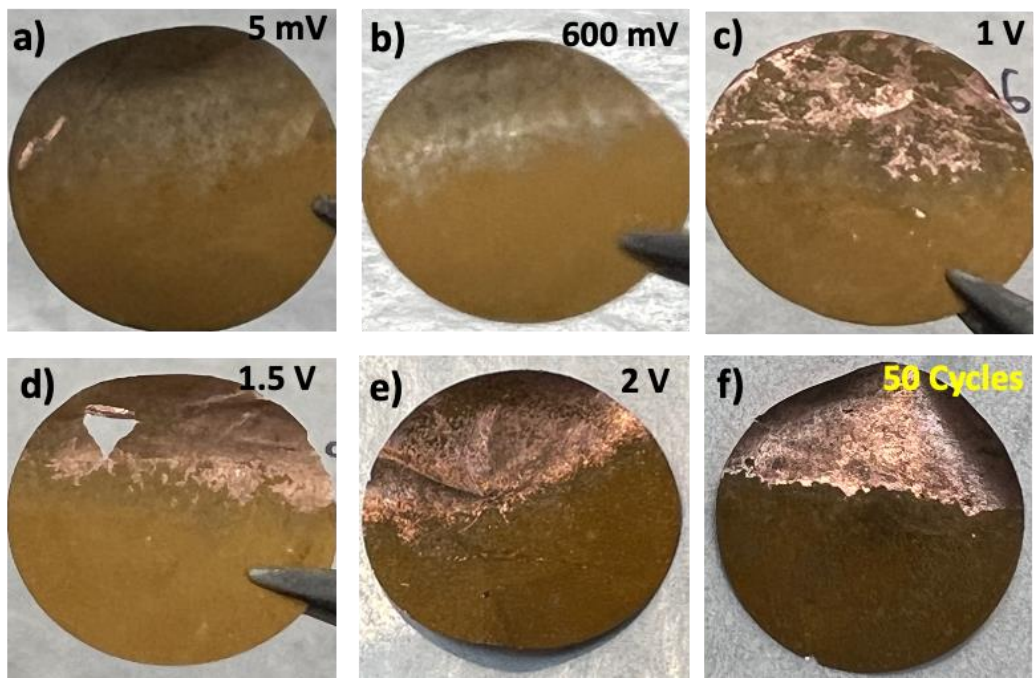

**Figure 4:** Binding ability studies (qualitative) on TiN-PI electrodes using scratch test during the 1st delithiation at a) 5 mV, b) 600 mV, c) 1V, d) 1.5 V, e) 2 V and f) after 50 cycles.

## 2.3. Projected consequences for full-cell testing

In most unaged Li-ion batteries, the anode will not experience potentials above 600-700 mV vs. Li/Li$^+$ when the full-cell is completely discharged.[17, 26] Since significant Li$^+$ extraction from the PI binder will not occur until ~800 mV vs. Li/Li$^+$ (Figure 1e), that implies that, in many cases, the polymer will never experience much delithiation in the full-cell. On the positive side, the binder retains its favorable adhesive properties (Figure 3). However, it also operates with an unreasonably low initial coulombic efficiency, as nearly all the reversible capacity of the PI remains trapped in the polymer.

To replicate the behavior of the polyimide in such conditions, a half-cell containing the TiN-PI electrode was cycled without a potentiostatic hold in between 5 mV and 600 mV (Figure 5). The PI exhibited a specific capacity of 1925 mAh/g during the first lithiation to 5 mV, as shown in Figure 5a. Further, when the cell is ramped to 600 mV, only 75 mAh/g of capacity could be extracted from the polymer. This corresponds to a first cycle coulombic efficiency of ~4%. From this difference of 1850 mAh/g, Figure 1b indicates that ~1000 mAh/g is the capacity that would otherwise be available should the electrode experience higher potential. Clearly, although PI binders suffer with irreversibility in the first cycle, most of this inefficiency actually stems from incomplete extraction of reversible capacity. The compounded effect of these losses is that it may be impractical to deploy cells using PI binder unless the electrodes are prelithiated.

Subsequent lithiation and delithiation resulted in capacity fade, with values saturating at 16.5 mAh/g after 50 cycles. Differential capacity plotted versus voltage indicates that the redox activity of PI occurs mostly around the voltage extremes (Figure 5b). These reversible capacities are accessed at markedly poor coulombic efficiencies (<93%).

During this study, we identified two additional mechanisms by which this PI binder could affect the capacity fade of full-cells. These arguments are general and likely applies to other high capacity, redox-active conductive binders. The first mechanism is a direct consequence of the conductivity and reactivity of the binder. In principle, the extent of SEI growth in the anode should be proportional to the surface area of electronically conductive materials that is accessible to the electrolyte. Since the PI binder itself becomes conductive when lithiated, it would also increase the active surface area of the electrode, contributing to additional SEI growth. This could create problems for calendar aging, as we demonstrate below.

The second mechanism by which this polyimide and similar binders can affect full-cell performance is more subtle and relates to the losses of binder capacity measured in half-cells (Figure 1c and Figure 5a). Capacity fade tends to lead to progressive increase in the potential experienced by the anode at the discharge of a full-cell.[17, 26] This is a consequence of the sloped voltage profile of most cathodes: lost $Li^+$ will lead to a lower $Li^+$ content in the cathode after discharge, increasing the terminal cathode potential; if the full-cell discharges to a constant voltage (which is the difference between electrode potentials), the anode potential must also increase. The consequence is that the more the cell ages, the more the PI will be delithiated during discharge of a full-cell. At first, it would seem that this additional delithiation would cause capacity fade. However, Figure 3a shows what happens during successive delithiation/lithiation half-cycles in a TiN-PI half-cell. Comparing the delithiation to 2 V with the ensuing lithiation (that is, discharge and charge, respectively, in the perspective of a full-cell), it is clear that the former exceeds the latter by a significant margin. In other words, more capacity is extracted in one cycle than the binder is capable to accommodate at the following half-cycle. That leaves some net amount of capacity that is returned to the cathode and no longer needed at the binder, effectively "prelithiating" the cell in situ. This intricate sequence of events can, thus, create the appearance that the cell is aging more slowly, at the expense of the binder. Examples of this mechanism are also shown below.

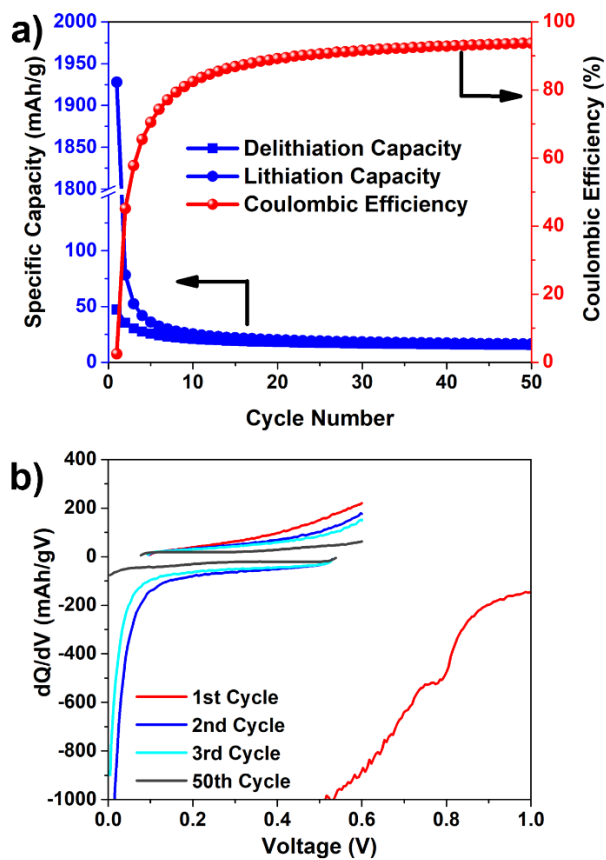

**Figure 5:** a) Electrochemical cycling of half-cells made using PI/TiN electrodes at C/5 rate with an upper cut-off of 600 mV and lower cut-off of 5 mV and its corresponding b) dQ/dV plot during the first three cycles.

Now, to the examples.

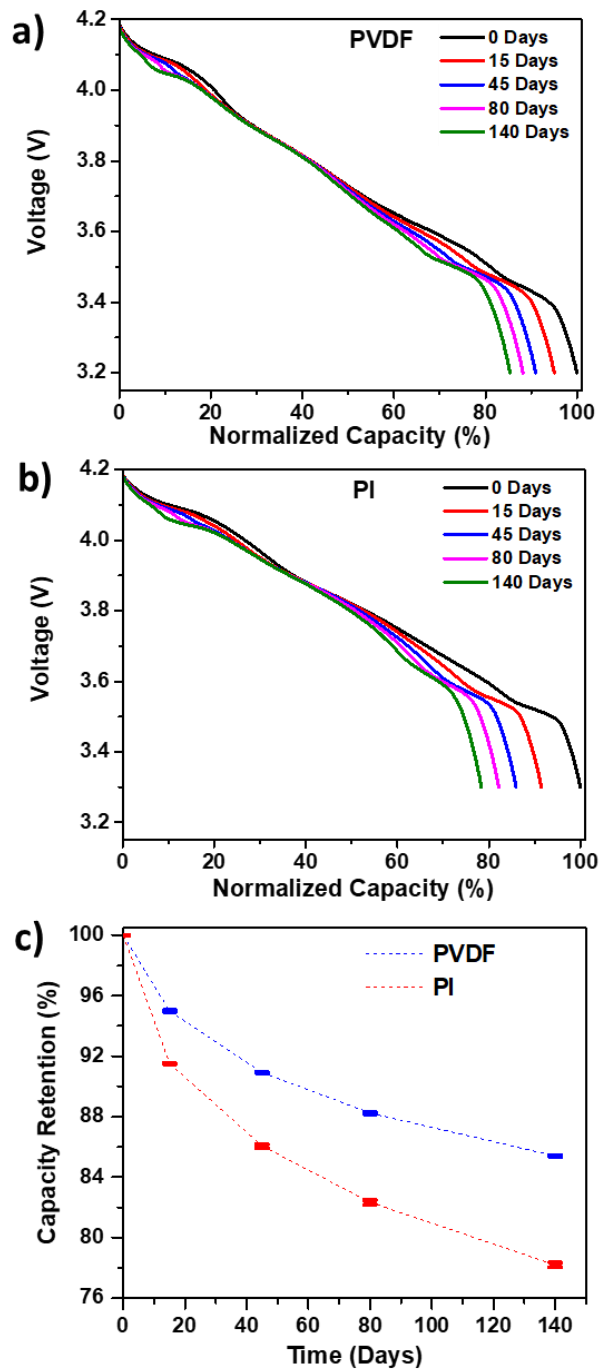

**Figure 6:** Calendar aging measurements of single layer pouch cells, made using NMC 811 cathode and a) graphite/PVDF, and b) graphite/PI anodes, at 45 °C for various durations. c) Standard deviation plot of capacity retention comparison with respect to the number of days of aging at 45 °C.

*2.4. Mechanism 1: the PI binder accelerates calendar aging.*

Calendar aging comprises the loss in performance that can be attributed to the extended exposure of the cell to specific storage conditions. This time-dependent degradation is particularly critical for Si anodes due to the dynamic nature of its SEI.[27] Since SEI formation requires electrons, and redox active binders have their own electron reservoir at the low potentials that the anode would experience in the charged state of a full-cell, the PI will also contribute to aging. To demonstrate this, calendar aging studies were carried out at 45 °C using single layer pouch cells with NMC811 and using two types of graphite anodes: one with PVDF and another with PI as binder. Although PIs are increasingly tested in silicon cells, graphite was the choice for these tests because of its stability that prevents any confusion between the anode and binder properties. The pouch cells were stored at 45 °C in the charged state (4.2 V) for extended periods, and capacity fade was tracked using occasional reference performance tests (RPTs) at 30 °C. Figure 6a, b shows representative discharge profiles for both types of cells; the profiles are taken from the second C/20 cycle of the respective RPT. After storage for 15, 45, 85 and 140 days, the PVDF cells lost about 5, 9, 12 and 15 % of capacity, respectively, while the PI cells lost 8.5, 14, 18 and 22% (Figure 6c). We credit these relatively large capacity fades, in part, to the absence of vinylene carbonate (VC) in the electrolyte.[28] In both cells, the rate of capacity fade was slightly higher for the initial 15 days of storage and later decreased during the subsequent aging. Clearly, losses in the PI cells are faster and larger than when an inert binder (PVDF) is used. The above study indicates that the calendar aging of the cell could be very much dependent on whether or not the binder is redox active. Development of novel polymer binders should include similar tests to demonstrate long-term stability.

*2.5. Mechanism 2: the PI binder can mask the capacity fade during aging.*

As discussed above, losses of binder capacity during delithiation can have a "prelithiation" effect on the cell, as more capacity is extracted from the PI than can be accepted by the polymer at the following half-cycle. This surplus of $Li^+$ can give the appearance of stability by offsetting some of the $Li^+$ that is lost to the SEI, to an extent that is proportional to the level of delithiation of the polyimide. Once more we used graphite anodes with PVDF or PI as binder, as the behavior of Si can be highly dependent of voltage cutoffs.[29] The full-cells included NMC811 as cathode. To vary the extent of delithiation of the polyimide, three discharge cutoffs were used: 2.5 V, 3.0

V and 3.2 V (PVDF) / 3.3 V (PI); the latter varied to enforce a similar anode potential despite differences in the initial inefficiency of each type of anode. At these cutoffs, the approximate anode potential at the end of discharge is >1.2 V, ~600 mV and <300 mV vs. Li/Li$^+$, respectively. At these potentials, the delithiation capacity of PI is large, small, and negligible (see Figure 1 and Figure 5), respectively. PVDF can be regarded as inert at this entire voltage range. The activity of PI becomes obvious comparing the tail-end of the discharge voltage profiles in Figure 7a, e. In the case of PVDF, where there is no redox contribution from the binder, the tail end of the voltage profile looks nearly vertical, while a slope is increasingly visible in PI cells at decreasing cutoffs.

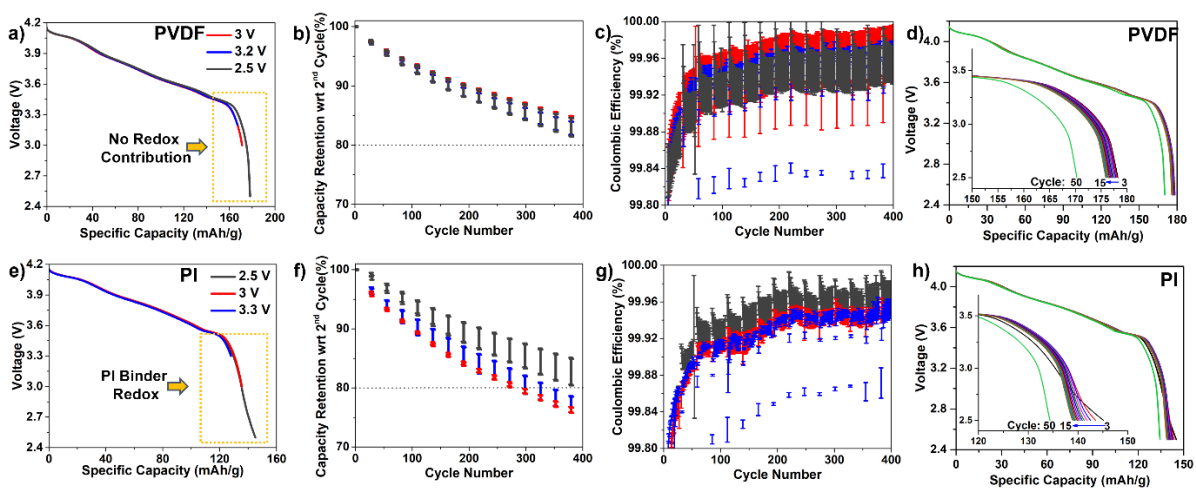

**Figure 7:** Galvanostatic cycling of cells made using NMC811 cathode and a-d) graphite-PVDF, and e-h) graphite-PI anodes, tested at varying lower cut-off voltages. a,e) Voltage profiles of the discharge curve at C/2 rate. b,f) Capacity retention during prolonged cycling measured during the RPT cycles to estimate the Li$^+$ inventory loss. c,g) Coulombic efficiencies during the C/2 cycling of the cells. d,h) Voltage profile at select cycles during C/2 discharge of cells going to 2.5 V lower cut-off voltage, with the inset showing the zoomed voltage profile at the tail end.

Also note the drastic difference in the specific capacity value obtained from the cells with each binder. Cells with PVDF binder exhibited >165 mAh/g capacity, whereas the cells with PI binder exhibited varying capacities depending on the cut-off voltage, with the maximum capacity around 145 mAh/g for the cell going to 2.5 V. This difference is due to the initial irreversibility of the PI binder.

The electrochemical cycling was carried out at C/2 rate, with RPTs at C/10 every 25 cycles. While C/2 aging was performed at all voltage cutoffs, RPTs were always done with the highest cutoff voltage (3.2 V for PVDF and 3.3 V for PI), enabling a direct comparison of the capacity retention across all testing conditions. As expected, PVDF cells displayed very consistent behavior irrespective of the cutoff voltages, presenting 83-85% retention after 400 cycles (Figure 7b) at an average coulombic efficiency of 99.93-99.94% (Figure 7c). The tail end of the profile of the cell going to 2.5 V does not vary upon cycling in the case of PVDF binder (Figure 7d). On the other hand, cells with the PI binder in the anode presented great dependence on the voltage cutoff. In this case, the cells going to 2.5 V (the ones experiencing reversible redox of the PI) displayed much better performance than the others, showing capacity retention of 82.7 % and coulombic efficiency of 99.94%, as opposed to 76-77% and 99.90%, respectively, for the other cells. When the PI is sufficiently delithiated during discharge, the performance is on par with the PVDF binder. The difference is that the similar metrics occur with varying levels of underlying stability: the "prelithiation" effect from delithiating the PI means that the cell can have a higher rate of SEI growth and yet exhibit capacity retention similar to PVDF cells. The tail end of the profile with PI binder shows drastic changes during cycling to 2.5 V (Figure 7h). The slope of the voltage profile fades during cycling and after 50 cycles, the profile is almost linear, similar to the one with PVDF (Figure 7d). This contrasting difference is caused by the drastic capacity fade of the PI binder upon cycling as observed in Figure 1 and Figure 5. Another important caveat is that, just as we demonstrated in Figure 4, this apparent cyclic stability will likely come at the expense of the adhesive properties of the binder. While this may not be an issue for graphite, it becomes extremely relevant considering the use of polyimide in silicon cells, which can undergo severe cyclic volume changes. We hope that this example serves as a cautionary tale of the detailed experimental planning that is required when judging the merits of new binder materials, as simply inspecting capacity retention can be insufficient to evaluate the true cell stability.

## 3. Conclusion

Several new polymeric binders with superior binding ability are being explored for Si anodes. Some of these binders undergo electrochemical doping/dedoping, which impart electrochemical redox activity and electronic conductivity. Such functions could be extremely helpful to improve cell performance, but could also destabilize the anode. The present work discusses the effect of utilizing such functional binders in Li-ion batteries through a model aromatic polyimide (PI) binder.

The electrochemical activity of the PI during the initial lithiation can be as high as 2200 mAh/g and the stable reversible capacity is close to 320 mAh/g. Solid-state NMR experiments revealed that the origin of this reversible capacity is the $\pi$-electrons in the aromatic groups of the PI. Although the PI exhibited high reversible capacity, the voltage profile presented large hysteresis that limits the delithiation of the PI when the anode experiences potentials below 600 mV vs. Li/Li$^+$. This value is close to the highest experienced by a conventional graphite anode in a full-cell, indicating that the PI tends to remain in the lithiated state in the cell. The lack of delithiation of the PI results in a poor coulombic efficiency of 4% (when considering the polymer alone) during the 1$^{st}$ cycle, indicating that pre-lithiation would be a necessary step for such binders irrespective of the active material being used. Furthermore, we showed that the adhesion of the binder decreases when the binder is delithiated (oxidized) above 1 V and continues to decrease considerably upon repeated cycling.

We also explored two mechanisms by which the PI can alter the electrochemical performance of model full-cells with standard NMC811 cathode and graphite anode: (i) the high electronic conductivity and redox behavior of the binder affecting its reactivity with electrolyte and SEI growth, and (ii) polymer degradation releasing Li$^+$ that affects the battery cycling. The reactivity between PI binder and electrolyte accelerated calendar aging, resulting in significant more capacity fade than when PVDF binder is used when cells are stored at 45 °C for 140 days. This shows how changes in the binder can have slow but deleterious effects on the stability of the cell. Additionally, we found that cycling of full-cells to various cut-off voltages showed variation in capacity retention depending on whether the PI binder was being oxidized or not. When the anode remained below 600 mV, the capacity retention of cells with PI binder was lower than when PVDF was used, but values became comparable when the PI was exposed to potentials > 1 V vs. Li/Li$^+$. This discrepancy was caused by an artificial increase in cell capacity at the expense of the binder, which appeared to degrade at high potentials, releasing excess Li$^+$ to the cathode that masks loss of capacity due to SEI growth. If such phenomenon remains undetected, cells could appear to be more stable than they actually are, precluding a correct assessment of aging trends. This work showed that despite the functional binders having several advantages over conventional binders like higher electronic conductivity and redox activity, it also comes with additional challenges like poor calendar life and electrochemical cycling stability that must be investigated in detail before employing them in full cells.

**Acknowledgements**

This research was supported by the U.S. Department of Energy's Vehicle Technologies Office under the Silicon Consortium Project, directed by Brian Cunningham, and managed by Anthony Burrell. The submitted manuscript has been created by UChicago Argonne, LLC, Operator of Argonne National Laboratory ("Argonne"). Argonne, a U.S. Department of Energy Office of Science laboratory, is operated under Contract No. DE-AC02-06CH11357. The U.S. Government retains for itself, and others acting on its behalf, a paid-up nonexclusive, irrevocable worldwide license in said article to reproduce, prepare derivative works, distribute copies to the public, and perform publicly and display publicly, by or on behalf of the Government.